\documentclass[iop,superscriptaddress]{emulateapj}
\usepackage{amsmath,epsfig,url,natbib}
\usepackage{graphicx,epstopdf,float,color,array}
\usepackage{rotating,multirow}
\submitted{}
\shorttitle{SOMfit}
\shortauthors{Shoubaneh Hemmati}

\definecolor{orange}{cmyk}{0,0.5,1,0}

\begin{document}
\title{Bringing manifold learning and dimensionality reduction to SED fitters}

\author{Shoubaneh Hemmati\altaffilmark{1}, 
Peter Capak\altaffilmark{2}, 
Milad Pourrahmani\altaffilmark{3}, 
Hooshang Nayyeri\altaffilmark{3}, 
Daniel Stern\altaffilmark{1},
Bahram Mobasher\altaffilmark{4}, 
Behnam Darvish\altaffilmark{5},
Iary Davidzon\altaffilmark{2},
Olivier Ilbert\altaffilmark{6},
Daniel Masters\altaffilmark{1},
Abtin Shahidi\altaffilmark{4}}
\email{shemmati@jpl.nasa.gov}
\altaffiltext{1}{Jet Propulsion Laboratory, California Institute of Technology, Pasadena, CA 91109, USA}
\altaffiltext{2}{IPAC, California Institute of Technology, 1200 East California Boulevard, Pasadena CA 91125, USA}
\altaffiltext{3}{University of California, Irvine, Irvine, CA 92697}
\altaffiltext{4}{University of California, Riverside, 900 University Ave, Riverside, CA 92521}
\altaffiltext{5}{California Institute of Technology, 1200 E California Blvd, Pasadena, CA 91125}
\altaffiltext{6}{Laboratoire D'astrophysique De Marseille, 38 Rue Frédéric Joliot Curie, 13013 Marseille, France}

\journalinfo{\textsuperscript{\textcopyright} 2019. All rights reserved. Submitted to the Astrophysical Journal Letters.}

\begin{abstract}

We show unsupervised machine learning techniques are a valuable tool for both visualizing and computationally accelerating the estimation of galaxy physical properties from photometric data. As a proof of concept, we use self organizing maps (SOMs) to visualize a spectral energy distribution (SED) model library in the observed photometry space.  The resulting visual maps allow for a better understanding of how the observed data maps to physical properties and to better optimize the model libraries for a given set of observational data.  Next, the SOMs are used to estimate the physical parameters of 14,000 $z\sim1$ galaxies in the COSMOS field and found to be in agreement with those measured with SED fitting.  However, the SOM method is able to estimate the full probability distribution functions for each galaxy up to $\sim 10^{6}$ times faster than direct model fitting. We conclude by discussing how this speed up and learning how the galaxy data manifold maps to physical parameter space and visualizing this mapping in lower dimensions helps overcome other challenges in galaxy formation and evolution.
\vspace{0.25cm}

\end{abstract}

\section{Introduction}

Understanding how galaxies evolve with time is one of the central questions in observational astronomy. But, the physical processes that drive galaxy evolution are a complicated combination of many physical processes that are intertwined.  For instance, gas accretion (e.g., \citealp{Rubin2012, Somerville2015, Zabl2019}) and its cooling into cold molecular gas that forms stars (e.g., \citealp{Saintonge2011, Ribaudo2011, Tacconi2013}), is disrupted by feedback from massive stars, AGNs and/or supernovae (e.g., \citealp{Springel2005, Fabian2012, Tombesi2015}).  Mergers of galaxies and other environmental processes further complicate this picture (e.g., \citealp{Moster2011, Hopkins2013, Fensch2017}). 

All of these processes are imprinted in the observed spectral energy distributions (SEDs) of galaxies.  The shape and normalization of these SEDs encodes physical properties such as the star formation rates and histories, stellar masses and dust content and distribution of galaxies. To estimate physical properties from the data a detailed physical model is used to synthesize observations for a wide range of possible formation scenarios (e.g. \citealp{Maraston2011,Conroy2013,Hayward2015}). This library of synthetic templates is then fitted to the data to determine the distribution of physical processes which might describe a given galaxy.  However, the complexity of the physics involved and the high-dimensionality of the physical parameter space mean understanding how the data and model space are related is very difficult.  Furthermore, the complexity of the physics means the model grids must be restricted to a controlled range of physical scenarios to make the fitting process computationally feasible.  

In \cite{Hemmati2018} and \cite{Masters2019}, we studied how Self Organizing Maps (SOMs, \citealp{SOM}) can be employed to improve redshift measurements. In this paper, we focus our attention towards the benefits of using a manifold learning algorithm (such as SOM) in the measurement of physical parameters of galaxies and present a framework for optimizing the parameters in theoretical models based on the data. In \S \ref{sec:data}, we describe the data and methods we adopt for the demonstrations in subsequent sections. \S \ref{sec:visual}, \S \ref{sec:pdfs}, and \S \ref{sec:speed} describe how we use the SOM to carry out parameter estimates and the distinct advantages compared to typical model fitting routines. Finally in \S \ref{sec:disc}, we discuss ways to build upon this machinery for future large surveys. Throughout this paper all magnitudes are expressed in AB system (\citealt{Oke1983}) and we use standard cosmology with $H_{0}=70 \ \rm km \ s^{-1}\  Mpc^{-1}$, $\Omega _{M}= 0.3$, and $\Omega _{\Lambda} = 0.7$.


\begin{figure*} 
\centering
  \includegraphics[trim=0cm 0cm 0cm 0cm, clip,width=1. \textwidth] {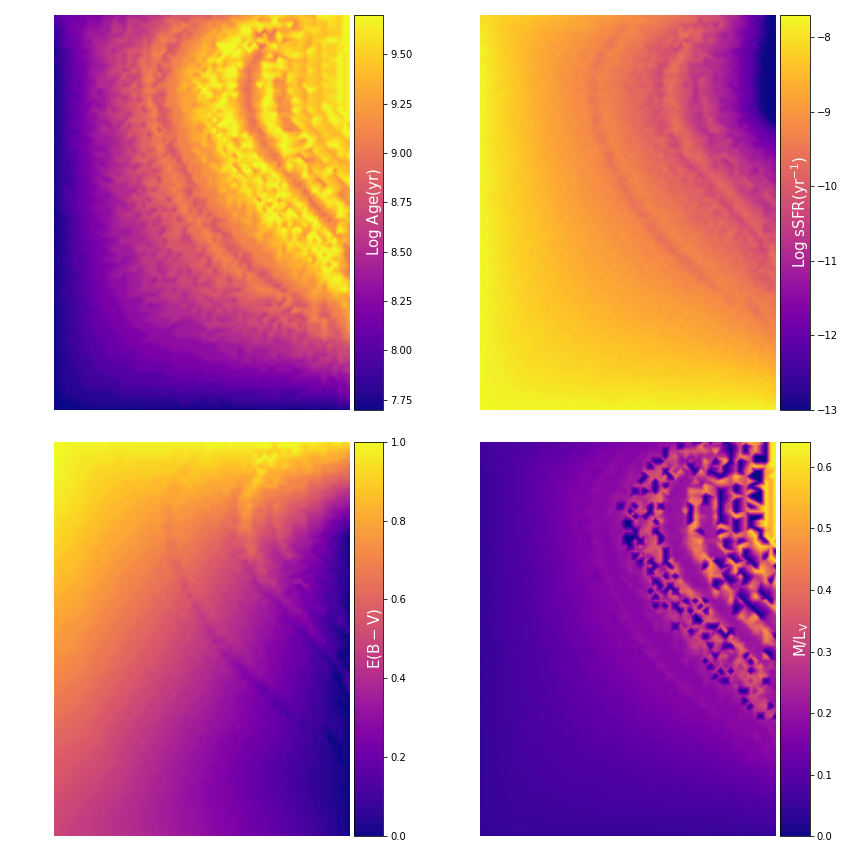}
\caption{Composite stellar population SEDs from our BC03 model library visualized on an $80\times 60$ SOM grid. The consecutive colors of the model SEDs in COSMOS optical and near-infrared broadband filter sets ($u-B,\ B-V,\ V-r,\ r-i^{+},\ i^{+}-z^{++},\ z^{++}-Y,\ Y-J,\ J-H,\ H-K_{s},\ K_{s}-ch1,\ ch1-ch2$) are used to train the SOM. The median of the physical properties of the models mapped to each cell are used to color the grids with stellar age, specific star formation rate (sSFR), dust reddening, and mass to light ratio.}
\label{fig:som_models}
\end{figure*}

\section{Data and Methodology}\label{sec:data}

The goal of this paper is to demonstrate how unsupervised machine learning can be used to understand and optimize model fitting.  To do that we set out a road map based on a demonstration data set.  Specific problems will require variations of the method described here.  In this section we begin by describing the data we used for the demonstration from the COSMOS survey in \ref{sec:cosmos} and the model library we have chosen to fit to the data in \ref{sec:models}.  We then give a brief overview of the SOM algorithm that we have chosen for this demonstration in \ref{sec:som} and note some alternative algorithms that could be used. 

\subsection{The Data \label{sec:cosmos}}

The COSMOS survey (\citealt{Scoville2007}) offers a suitable data sets for studying galaxy formation and evolution, providing both area ($\sim 2\rm deg^{2}$) and depth ($\sim 26$ AB mag in the optical) across a broad range of wavelengths with around two million detected galaxies. We use the broad-band photometric data from the latest publicly available multi-wavelength catalog in the COSMOS (see \citealt{Laigle2016} for details), namely the CFHT/Megacam $u^{*}$, Subaru/Suprime-CAM $B,V,r,i^{+},z^{++}$, VIRCAM/UltraVISTA $Y,J,H,K_{s}$, and \textit{Spitzer}/IRAC $ch1,ch2$. In this paper, we do not include longer wavelength observations.  For demonstration purposes we also fix redshifts to those reported by the COSMOS team and restrict all our analysis to 13,781 galaxies in the redshift range $0.8<z<1.2$.

\subsection{Spectral Energy Distribution model library \label{sec:models}}

We chose the commonly used BC03 simple stellar population synthesis template library (\citealt{Bruzual2003}) restricted to $z\sim1$ for our demonstration.   While not optimal, this simple library is adequate for demonstration purposes. A forthcoming paper, Davidzon et al. (in prep.), applies these techniques using more sophisticated Hydrodynamical simulations across all redshifts from the Horizon AGN simulation (\citealt{Laigle2019}). 
 
Here we adopt a Chabrier initial mass function (\citealt{Chabrier2003}), exponentially declining star formation histories ($e^{-t/\tau}$ with $\tau$ values in the range $8.5-10$), a sub-solar metallicity ($0.4\ Z_{\odot}$), a range of stellar ages ($7.7<log(age/yr)<10.0$), and a \cite{Calzetti2000} extinction law with a range of reddening values ($0<E(B-V)<1$). To generate synthetic photometry that can be fit to our data we redshifted the models and integrated them against the same filter transmissions used for COSMOS observations (see \S \ref{sec:cosmos}). This yields 13,776 models with the same parameters as the observed data.

\subsection{Unsupervised learning algorithm \label{sec:som}}

We chose to use the Self Organizing Map (SOM) algorithm for this demonstration because it provides an easily understood visual map of the high-dimensional data along with a vector representation of the model space.  Furthermore, we have used it in previous papers, \cite{Hemmati2018} and \cite{Masters2019}, for related problems so contemporary readers would be familiar with the method. We note that other manifold learning and dimensionality reduction methods such as but not limited to the t-SNE (t-Distributed Stochastic Neighbor Embedding; \citealp{TSNE}), UMAP (Uniform Manifold Adaptation and Projection; \citealp{UMAP}), GNG (Growing Neural Gas networks; \citealp{GNG}) can also be employed and may be more optimal for some problems.

SOMs are a class of unsupervised neural networks which reduce the dimensionality of data, while preserving the topology. Therefore,  neighbouring points in the N-dimensional space are going to be neighbours in the 2-D projected map. Each neuron corresponds to a pixel on the map as well as a vector into the photometric data space.  The neurons/vectors can be linearly combined to represent any object in the data set.  Alternatively, the neurons can be thought of as bins in the high-dimensional data space.  Data points which are statistically similar will get assigned to the same bin, reducing the number of point required to describe the data set. 

In previous works (e.g., \citealp{M15,Hemmati2018,Masters2019}) we demonstrated how the manifold of galaxy colors can be learned using a SOM.  We then showed a SOM can be used to understand and calibrate photometric redshifts as well as understand variations in galaxy spectral types with color.  We refer the interested reader to those work for more details about the methodology. 

For this work we use the {\sc sompy} package, a \texttt{python} library for self organizing maps and adopt an $80\times 60$ rectangular grid. The size is chosen to represent the data distribution well (\citealp{Hemmati2018}). The consecutive colors of galaxies as measured in the filters listed in \S 2.1 define the multi-dimensional (11D) color-space to be learned. Hence, each cell on the constructed grid has an eleven dimensional weight vector whose values  (initially assigned by principle component analysis) change until they fully represent the training data. 

\section{Visualizing and Optimizing model libraries}\label{sec:visual}

\begin{figure} 
\centering
  \includegraphics[trim=0cm 0cm 0cm 0cm, clip,width=0.5
  \textwidth] {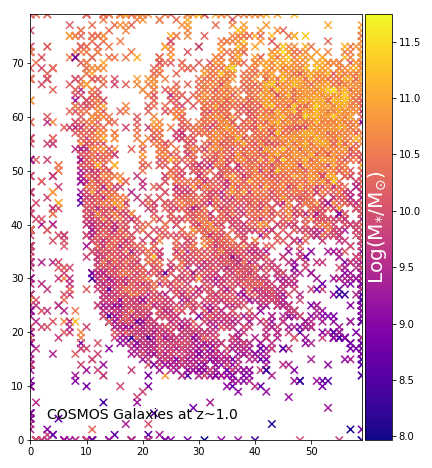}
\caption{COSMOS galaxies at $z\sim 1$ are mapped to the SOM trained with theoretical models. Empty regions of the SOM represent combinations of the theoretical parameter space that do not have observed counterparts. Objects at the edge of the map typically are poorly fit by the model.  This visualization enables the optimization of model libraries based on information from the observed data space. Galaxies in the map are color-coded with their estimated stellar masses from \citet{Laigle2016}.}
\label{fig:dataonSOM}
\end{figure}

Having a large number of physical parameters (dimensions) in a model SED library is valuable for understanding the galaxy population.  However with large numbers of parameters traditional model fitting analysis becomes computationally unfeasible even with advanced techniques such as Monte-Carlo-Markov-Chain (MCMC) fitting (e.g., \citealt{Johnson2013,2016MNRAS.461.3432S,Wilkinson2017}).  Furthermore, understanding how each parameter is shaping the overall fit, which ones are correlated, and what is actually constrained by the data becomes very challenging. 

\begin{figure} 
\centering
  \includegraphics[trim=0cm 0cm 0cm 0cm, clip,width=0.53 \textwidth] {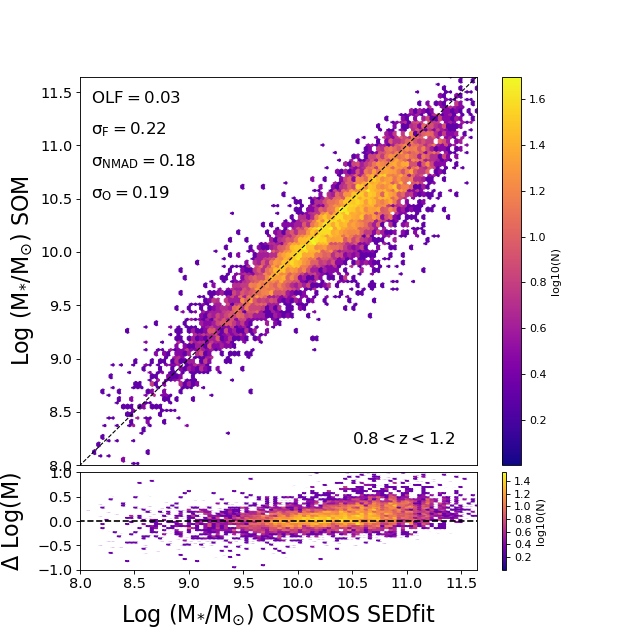}
\caption{Stellar masses of COSMOS $z\sim 1$ galaxies measured by mapping them to the SOM trained with BC03 models (SOMfit) compared to stellar masses from the \citep{Laigle2016} catalog measured with LePhare. The differences are well within the expectations from using different model libraries.}
\label{fig:Masscomp}
\end{figure}

\begin{figure*}
\centering
  \includegraphics[trim=0cm 0cm 0cm 0cm, clip,width=1.0 \textwidth] {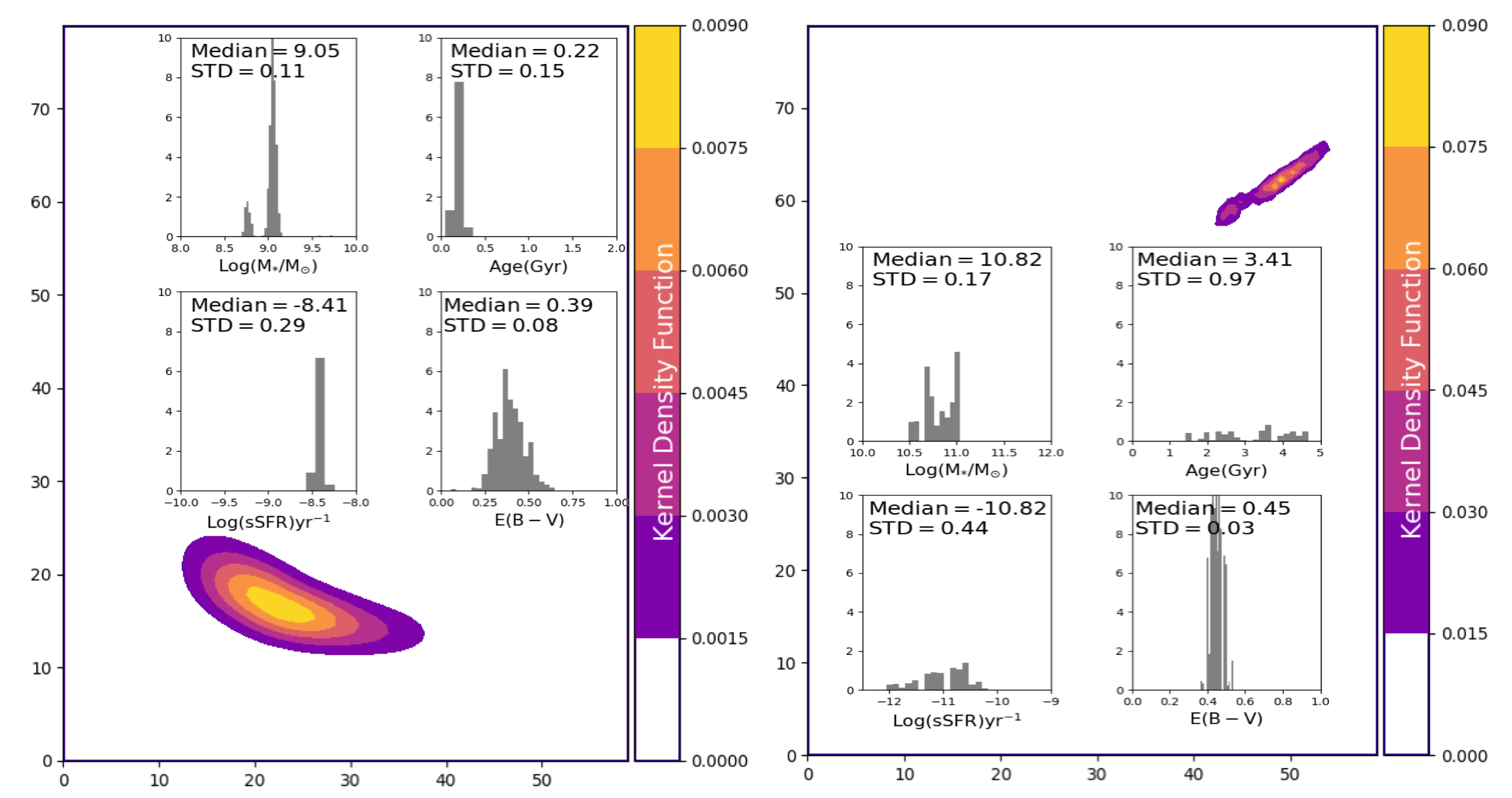}
\caption{To quantify the uncertainties in the measurement of physical parameters with the SOM, many realizations (here 1000) of two sample galaxy SEDs perturbed within the photometric errors in each band are mapped to the SOM to generate the Kernel Density Function. The inset plots represent the confidence on each of the measured physical parameters (top-left to bottom-right panels: stellar mass, stellar age, sSFR and E(B-V)).}
\label{fig:uncertainty}
\end{figure*}

A primary advantage of using unsupervised dimensionality reduction methods like the SOMs is the ability to visualize complex data spaces. In the model space, the dimensionality increases with the number of physical properties considered, so varying the IMF or adding different dust properties would increase the dimensionality.  In this paper we focus on visualizing the models in data space, but note that the same techniques could be used to visualize in the model parameter space.  For photometric data, the dimensions of the observed parameter space increase with increased number of colors (filters).  In other words, increasing the wavelength range by adding broad band filters or the resolution by adding intermediate band filters would increase the dimensionlity of the data space. 

We begin by taking the synthetic model photometry described in \S \ref{sec:models} and creating a SOM that represents those synthetic photometry data.  In Figure \ref{fig:som_models} we visualize model parameters in photometric data space by coloring each pixel in the SOM with its median physical parameter from the models.  The smooth areas of the maps represent regions where the data space directly maps to physical parameters, so the data can strongly constrain the physical parameter in question. The more stochastic parts of the map indicate regions where the data are not constraining of a given physical parameter, so the same data can describe multiple possible physical properties. It is clear from looking at these maps that there is a region of data space where age and Mass-to-Light ratios are poorly constrained by the data and other regions where the properties are well constrained. 

The next question is which portions of the model space are occupied by real galaxies and do the galaxies occupy smooth or stochastic parts of the model space. In Figure \ref{fig:dataonSOM} we map the data for COSMOS $z\sim 1$ from \S \ref{sec:cosmos} galaxies to the model SOM. The empty regions in the map indicate regions where no real object corresponds to those model parameters.  For instance, the large empty region in the bottom right of the plot indicates that a combination of very young age, low extinction and high star formation rate per unit mass (specific star formation rate, sSFR) is not observed in COSMOS at $z\sim 1$.  Furthermore, the points at the very edge of the plot are also at the edge of the model space and hence are not well fit.  The presence of these points indicates galaxies that are not well described by the models and that additional parameters, such as the presence of an Active Galactic Nuclei (AGN) may be needed.

This or similar analysis can be used to optimize the model library.  Regions of the model space that are not occupied by real objects should be removed or down weighted based on the data statistics. Including them in a fitting library with a flat prior will lead to biased physical property estimates because noise in the data will scatter objects into regions of model space that do not physically exist.  Furthermore, the expansion of the model library could be guided by the points that are not well described by the model library. 

\section{Optimizing model fitting}\label{sec:pdfs}

Mapping an observed galaxy to the SOM trained on the model libraries is mathematically equivalent to finding the best-fit template in a model library. However, because the SOM optimally represents the data space, observationally indistinguishable models have already been assigned to the same SOM vector.  This methodology effectively "compresses" the computation by pre-grouping models that will have indistinguishable fits to the data. So, rather than fitting every model one only needs to fit each point in the SOM to estimate the posterior likelihood distribution.   

In our simple demonstration (see Figure \ref{fig:uncertainty}) the likelihood distribution in physical parameter space is calculated by assigning the median property of the models at each SOM point.  We then weight each SOM point by its likelihood with respect to the data in the same way as if one were fitting a model library. We note that a better weighting scheme could easily be devised, but is beyond the scope of the demonstration here.  

In Figure \ref{fig:Masscomp}, we compare best fit stellar mass measured from the SOM to that from the COSMOS catalog estimated using the LePhare SED fitting code (\citealp{Arnouts1999,Ilbert2006}). In this comparison, we use statistics similar to those defined in \cite{Nayyeri2017}, where $\sigma_{\rm F} = rms(log(M_{\rm SOM})-log(M_{\rm COSMOS}))$, $\sigma_{\rm NMAD}=1.48\times median(|log(M_{\rm SOM})-log(M_{\rm COSMOS})|)$ and $\sigma_{\rm O}= rms(log(M_{\rm SOM})-log(M_{\rm COSMOS}))$ after removing the outliers with the outlier fraction(OLF) defined as the fraction of
objects with $\Delta log(M)=(log(M_{\rm SOM})-log(M_{\rm COSMOS}))>0.5$. The scatter of $\sim 0.2$ and $3\%$ OLF is comparable with the values reported in the literature when various SED fitting codes are compared (e.g., \citealp{Mobasher2015, Nayyeri2017}) and we note more advanced models including emission lines and other features (i.e., delayed SFH and two attenuation curves) were included in the COSMOS results.  Furthermore, using a more thorough weighting scheme and including redshift errors would likely further improve results. 

Another compelling feature of the SOM is that the errors on the physical properties are quantified. The method naturally builds up a probability distribution function (PDF) for each physical parameter based one the likelihoods of the models. For instance, one can perturb the photometric error in each broad-band filter to have many realizations of a galaxy SED and mapping them to the SOM. In Figure \ref{fig:uncertainty} we show the kernel density function of 1000 realizations of the SED, drawn from the estimated errors, mapped to the SOM with contours for two sample galaxies. The corresponding distribution of physical properties are plotted in the inset panels. 

It is clear in this visualization how the error contour for a given galaxy spreads over the SOM grid and how that is related to the physical parameters shown in Figure \ref{fig:som_models}.  

 By comparing Figure \ref{fig:uncertainty}, and \ref{fig:som_models} it is clear how parameters are co-variant with error.  In many SED fitting routines, all theoretical model SEDs have the same probability of being fitted to a set of observations. A comparison of Figure \ref{fig:dataonSOM} and \ref{fig:som_models} shows that photometric errors will create probability in non-physical regions of the model space where no objects occur.  So the use of a flat prior creates a clearly biased result. A better approach would be to use the density of data on the SOM grid as a weight on the model fit. As demonstrated in \citet{Masters2019} this results in a less biased result while still preserving the ability to fit legitimately rare sources. 

\section{Computational Speed}\label{sec:speed}

Measuring galaxy physical properties using SOMs is also computationally efficient. Once the SOM is trained with a library of model SEDs, there is no need to keep the large number of models in the memory and any set of observations can be mapped to the trained SOM rapidly.  In addition, the size of the SOM is determined by the data space, not the model space, so the required computational time to fit a set of models is largely invariant to the complexity and size of model grids and is instead determined by the dimensionality of the data set.  So its possible to fit multiple very complex models to the data.  This is a significant advantage for future large area surveys with billions of galaxies such as Euclid, LSST, and WFIRST because it will allow for rapid re-fitting of multiple SED model libraries.    

\begin{figure} 
\centering
  \includegraphics[trim=0cm 0cm 0cm 0cm, clip,width=0.49 \textwidth] {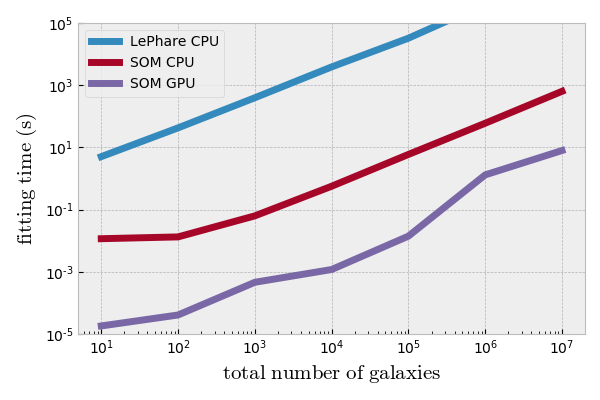}
\caption{Computational speed of mapping different numbers of galaxies to the SOM (on a CPU and on a GPU) compared to SED fitting with LePhare. Using our mapping we increase the performance by $\sim 10^{3-4}$ on a CPU and $\sim 10^{5-6}$ on a GPU compared to SED fitting.}
\label{fig:speed}
\end{figure}

Furthermore, the mapping of galaxies to the SOM is inherently parallel and hence is easily applicable on a GPU or a large cluster.  To verify the speed up we converted the mapping phase (i.e. finding the best matching unit of the SOM with least $\chi^{2}$) to a vectorized GPU friendly code using \textit{PyTorch}(\citealt{pytorch}) and \textit{Numba} (\citealt{Numba}). Figure \ref{fig:speed} shows the processing time of measuring physical properties of galaxies using the LePhare SED fitting code (REF) compared to mapping the galaxies on to the SOM using both a CPU on a Core-i7 machine and an \textit{NVidia Titan V GPU}. The SOM is a $\sim3-4$ orders of magnitude faster than LePhare on a CPU and $\sim 5-6$ orders of magnitude faster on a GPU.  This means even a single threaded CPU could measure the physical parameters of over a million galaxies per minute and a single GPU could re-compute stellar masses for the 30 billion objects expected from LSST in under nine hours compared with several years for LePhare. So even a modest cluster of GPU machines could re-compute these parameters for all of LSST interactively.   

\section{Discussion}\label{sec:disc}
We present the idea of using manifold learning as an augmentation to traditional SED fitting.  We present the SOM method in particular as a way of visualizing and optimizing a library of theoretical models.  We then show the same method can be used for fast estimation of galaxy physical properties and how the data maps to these properties. 

In previous works we showed the manifold mapping approach improves  photometric redshifts (e.g., \citealp{M15,Masters2017, Masters2019,Hemmati2018}). In this work, as a proof of concept, we constrained our analysis to only a small redshift bin and a simple library of model SEDs to clearly demonstrate the method.  We show the visualization and computational improvements have clear utility for other studies. 

Beyond fitting objects the techniques we present here have multiple uses.  Most popular galaxy classifications in extragalactic astronomy use only a few of broadband colors to select distinct classes of galaxies such as the Lyman break, Balmer break, or the "UVJ" selections (e.g., \citealp{Steidel1999, Nayyeri2014, Williams2009}). 
Manifold mapping techniques can offer an excellent alternative tool that uses all of the color information.  For instance, the SOM, with its preserved topology could be used to specify N-dimensional color selections for specific classes of galaxies.  The exact cuts could be defined using the visualization presented in Figure \ref{fig:som_models}.

The speed of this method also allows for new possibilities.  For instance, it could be applied to entire multi-color data sets at the pixel or resolution element level.  This would enable exploration of resolved kpc-scale distribution of physical properties (e.g., \citealp{Wuytys2012, Hemmati2014, Hemmati2015}) or definition of object de-blending based on these maps. Finally, other observational data besides colors could be included in this method to create multi-measurement studies or fits.  For instance, environmental measures or morphology could be included. We are now extending this work to more complex models based on hydronamical simulations (Davidzon et al- in prep.) and use a similar technique to predict higher resolution spectral features like emission line strengths from broadband information (Hemmati et al - in prep.).

This work used {\sc SOMPY}, a python package for self organizing maps (main contributers: Vahid Moosavi @sevamoo, Sebastian Packmann @sebastiandev, Iv\'an Vall\'as @ivallesp). We are thankful to NVIDIA for the GPU granted as their academic grant program. Parts of this research were carried out at the Jet Propulsion Laboratory, California Institute of Technology, under a contract with the National Aeronautics and Space Administration.

\bibliography{somfit.bib}

\end{document}